# High thermoelectric performance can be achieved in black phosphorus

J. Zhang, H. J. Liu[*], L. Cheng, J. Wei, J. H. Liang, D. D. Fan, P. H. Jiang, L. Sun, J. Shi

*Key Laboratory of Artificial Micro- and Nano-Structures of Ministry of Education and School of Physics and Technology, Wuhan University, Wuhan 430072, China*

Few-layer black phosphorus has recently emerged as a promising candidate for novel electronic and optoelectronic device. Here we demonstrate by first-principles calculations and Boltzmann theory that, black phosphorus could also have potential thermoelectric applications and a fair *ZT* value of 1.1 can be achieved at elevated temperature. Moreover, such value can be further increased to 5.4 by substituting P atom with Sb atom, giving nominal formula of $P_{0.75}Sb_{0.25}$. Our theoretical work suggests that high thermoelectric performance can be achieved without using complicated crystal structure or seeking for low-dimensional systems.

## 1. Introduction

As a clean and viable way to solve the global energy and environmental crisis, thermoelectric energy conversion that is capable of creating electricity from heat and vice versa has received extensive interest over the past few decades. The efficiency of a thermoelectric material is determined by the dimensionless figure of merit:

$$ZT = \frac{S^2\sigma}{\kappa_e + \kappa_p}T, \quad (1)$$

where $S$ is the Seebeck coefficient, $\sigma$ is the electrical conductivity, $T$ is the absolute temperature, and $\kappa_e$ and $\kappa_p$ are the electronic and lattice thermal conductivities, respectively. Intuitively, good thermoelectric material should possess a large Seebeck coefficient, a high electrical conductivity, and a low thermal conductivity. However, these transport parameters are closely interrelated and related to the electronic and crystal structure, as well as the carrier mobility and concentration. As a result, the *ZT* value of some good thermoelectric materials such as $Bi_2Te_3$ has remained at about 1.0 for several decades, whereas a ZT ~ 3 is necessary to compete with conventional refrigerators or power generators. To significantly enhance the *ZT* value, much efforts have been devoted to search for new bulk materials with multi-components and/or complicated crystal structures, such as AgPb$_m$SbTe$_{2+m}$ [1], Bi$_{0.5}$Sb$_{1.5}$Te$_3$ alloys [2], and PbTe-SrTe [3]. On the other hand, improved thermoelectric performance has also been found in many low-dimensional systems, such as Bi$_2$Te$_3$/Sb$_2$Te$_3$ superlattice structure [4], PbSeTe/PbTe quantum-dot superlattice [5], and Si nanowires [6].

---

[*] Author to whom correspondence should be addressed. Electronic mail: phlhj@whu.edu.cn.

As a single-component material which is constituted by the low-cost and earth-abundant element, the possibility of using black phosphorus (BP) as thermoelectric materials has recently attracted growing interest. For example, Lv *et al.* found that the power factor (PF=$S^2\sigma$) of BP can reach as high as 118.4 μWcm$^{-1}$K$^{-2}$ at appropriate carrier concentration [7]. However, they predicted that the thermoelectric performance of BP is poor at room temperature, which may be caused by relatively large lattice thermal conductivity (12.1 W/mK, taken from that of polycrystalline BP) [8]. Qin *et al.* showed that the maximal *ZT* is 0.72 at 800 K and can be enhanced to 0.87 by a proper strain [9]. Flores *et al.* experimentally measured the thermoelectric properties of BP and found that it is a *p*-type semiconductor with a Seebeck coefficient of 335 μV/K at room temperature. Besides, the power factor of BP at 385 K is 2.7 times higher than that at room temperature, which is caused by the reduced electrical resistance with increasing temperature [10]. Although these works show that BP could be a potential intermediate temperature thermoelectric material, the predicted *ZT* value is still not comparable with that of conventional thermoelectric materials. In this work, using first-principles calculations and Boltzmann transport theory, we provide a systematic investigation on the thermoelectric properties of BP within the reasonable carrier concentration range. We demonstrate that BP exhibit a fair *ZT* value at elevated temperature even with a relatively higher lattice thermal conductivity. More importantly, we show that the substitution of P atom with Sb atom can not only significantly reduce the lattice thermal conductivity of BP, but also increase the density of states (DOS) around the Fermi level. As a result, the *ZT* value of BP can be optimized to as high as 5.4 at 800 K, giving strong evidence that high thermoelectric performance can be achieved without using complicated crystal structure or seeking for low-dimensional systems.

## 2. Methodology

### 2.1 Electronic structure calculations

The geometry optimization and band structure calculation of BP are performed by using the first-principles plane-wave pseudopotential formulation [11, 12, 13] as implemented in the Vienna *ab-initio* (VASP) [14] code. The exchange correlation functional is employed in the form of Perdew-Burke-Ernzerhof (PBE) [15] with the generalized gradient approximation (GGA). The cutoff energy for the plane wave basis is set to be 500 eV. In the ionic relaxation and charge density calculation, Monkhorst-Pack ***k***-mesh [16] of 8×8×10 is used. For the optimization of crystal structure, van der Waals (vdW) interaction is considered at the vdW-DF level with optB88 exchange functional (optB88-vdW) [17, 18]. The atom positions are fully relaxed until the magnitude of force acting on all atoms become less than 0.01 eV/Å. For the calculation of electronic structures, we use the modified Becke-Johnson (mBJ) [19, 20] exchange potential, which can give accurate band gaps for many semiconductors.

### 2.2 Boltzmann transport theory for electron

For the electronic transport, we apply the semiclassical Boltzmann transport equation [21] with the relaxation time approximation. Such approach has been successfully used to predict the transport coefficients of some known thermoelectric materials, and the theoretical calculations are found to agree well with the experimental results [22, 23, 24, 25]. In terms of the so-called transport distribution function $\Xi(\varepsilon) = \sum_{\vec{k}} \vec{v}_{\vec{k}} \vec{v}_{\vec{k}} \tau_{\vec{k}}$, the Seebeck coefficient ($S$) and the electrical conductivity ($\sigma$) can be expressed as:

$$S = \frac{ek_B}{\sigma} \int d\varepsilon \left( -\frac{\partial f_0}{\partial \varepsilon} \right) \Xi(\varepsilon) \frac{\varepsilon - \mu}{k_B T}, \quad (2)$$

$$\sigma = e^2 \int d\varepsilon \left( -\frac{\partial f_0}{\partial \varepsilon} \right) \Xi(\varepsilon), \quad (3)$$

where $\vec{v}_{\vec{k}}$ and $\tau_{\vec{k}}$ are the group velocity and relaxation time at state $\vec{k}$, respectively. The other parameters are the equilibrium Fermi function $f_0$, the Boltzmann constant $k_B$, and the chemical potential $\mu$. The optimal carrier concentration is obtained by integrating the DOS from the desired chemical potential to the Fermi level ($\mu = 0$). The carrier relaxation time can be extracted from the experimentally measured carrier mobility ($\mu$) and effective mass ($m^*$) [26]. Details are given in the Supplement Information. In order to obtain reliable transport coefficients, we use fine *k*-mesh of 26×26×32 and 18×18×22 to calculate the band energies of pristine and substituted BP, respectively.

The electronic thermal conductivity $\kappa_e$ is calculated by using Wiedemann-Franz law [27]:

$$\kappa_e = L\sigma T, \quad (4)$$

where $L$ is the Lorenz number. For three-dimensional materials, depending on the reduced Fermi level $\xi = E_f / k_B T$, the Lorenz number can be expressed by [28, 29]:

$$L = \frac{\kappa_e}{\sigma T} = \left( \frac{k_B}{e} \right)^2 \left[ \frac{7 F_{5/2}(\xi)}{3 F_{1/2}(\xi)} - \left( \frac{5 F_{3/2}(\xi)}{3 F_{1/2}(\xi)} \right)^2 \right], \quad (5)$$

where $F_i(\xi) = \int_0^\infty \frac{x^i dx}{e^{x-\xi}+1}$.

**2.3 Boltzmann transport theory for phonon**

The lattice thermal conductivity of BP can be calculated by using phonon Boltzmann transport equation (BTE) with relaxation time approximation as implemented in the ShengBTE code [30, 31, 32]. In this approach, the thermal conductivity along the $\alpha$ direction can be calculated by:

$$\kappa_\alpha = \frac{1}{N_q V} \sum_{\vec{q},j} C_{\vec{q},j} v^2_{\vec{q},j,\alpha} \tau_{\vec{q},j}, \tag{6}$$

where $C_{\vec{q},j}$ is the specific heat contribution of the phonon mode with the wave vector $\vec{q}$ and polarization $j$, $v_{\vec{q},j,\alpha}$ is the phonon group velocity, and $\tau_{\vec{q},j}$ is the phonon relaxation time. $N_q$ is the number of sampled *q* points in the Brillouin zone and $V$ is the volume of the unit cell.

During the thermal conductivity calculations, the only inputs are the second order and third order force constant matrix, which can be extracted from the first-principles calculations. Here the exchange-correlation functional is the same as that done for the electronic structure calculations. A 4×4×5 supercell is adopted to calculate the second-order and third-order force constant. The interactions up to the fourth nearest neighbors are considered when dealing with the anharmonic one. The phonon BTE method has already been used to estimate the lattice thermal conductivity of phosphorene [33, 34, 35] and phosphorene nanoribbons [36].

## 3. Results and discussion
### 3.1 Lattice structure and electronic properties of BP

The crystal structure of bulk BP is shown in Figure 1, which exhibit an orthorhombic symmetry and has the space group of *Cmca*. In the *xz* plane, each P atom is covalently bonded with three nearest neighboring atoms to form a puckered structure. Along the *y* direction, the neighboring layers are held together by weak van der Waals forces. For the convenience of discussions, we define the zigzag, the out-of-plane, and the armchair direction as *x*, *y*, and *z* direction, respectively. The predicted structural parameters of BP by using different kinds of exchange and/or correlation functionals are presented in Table I. We see that the lattice constants calculated with van der Waals corrections explicitly included by using DFT-D2 [37] and optB86b-vdW functional [17] are very close to the experimental values [38].

However, the electronic band structure based on these "best fit" lattice constants gives a dissatisfied band gap, as indicated in Figure S1 of the Supplement Information. As an alternative, we choose the optB88-vdW functional [17, 18] with mBJ exchange potential [19, 20] to obtain an accurate band structure. Figure 2 shows the calculated band structure and DOS of BP. We see that BP is a semiconductor with a direct band gap of 0.31 eV at the Z point, which agrees well with that found previously [39, 40, 41]. Moreover, we find that the highest valence band and the lowest conduction band disperse significantly along the Q-Z direction (*z* direction or armchair direction) and Γ-Z direction (*y* direction or out-of-plane direction), which indicates very small effective mass of electrons and holes. Table II lists the calculated carrier effective masses along the *x*, *y*, and *z* directions, where we find that the one along the *z* direction is the smallest, especially for the case of holes. Such small effective mass would set a crucial precondition for BP to exhibit a high carrier mobility, which is confirmed by the experimental results [42] and suggests that the carrier transport of BP along the *z* direction is quite favorable and may be beneficial to its thermoelectric performance. On the other hand, we see the DOS of BP at the bottom of conduction band exhibits a sharper peak than that at the top of valance band. Since the magnitude of Seebeck coefficient is determined by the DOS distributed around the Fermi level [43], we expect that the *n*-type BP may have a higher Seebeck coefficient than *p*-type system, as will be discussed in the following.

**3.2 Electronic transport properties**

We now move to the investigation of electronic transport properties. Figure 3 plots the calculated transport coefficients along the *x*, *y*, and *z* directions as a function of carrier concentration from $1\times10^{19}$ to $1\times10^{21}$ cm$^{-3}$. We see that at both 300 K (Fig. 3(a)) and 800 K (Fig. 3(b)), the absolute value of Seebeck coefficient ($S$) of *n*-type BP along the *x* direction is always the highest among the three directions. However, this is not the case for the *p*-type system where the Seebeck coefficients along the three directions are almost identical to each other in the carrier concentration range considered. Moreover, for all the three directions, we find the absolute value of Seebeck coefficient of *n*-type system is larger than that of *p*-type system at the same carrier concentration. This is consistent with the sharper DOS at the bottom of the conduction band, as observed in Fig. 2(b). If we focus on the high temperature case, we see from Fig. 3(b) that the absolute values of Seebeck coefficients for all the three directions are always larger than those found at 300 K. Moreover, a maximum Seebeck coefficient of 392 μV/K can be achieved along the *x* direction at a moderate carrier concentration of $1\times10^{19}$ cm$^{-3}$. Such value is much larger than those of conventional thermoelectric materials and suggests that BP could have very favorable thermoelectric performance at elevated temperature.

Fig. 3(c) shows the room temperature electrical conductivity ($\sigma$) of BP as a function of carrier concentration. We see that for all the three directions the electrical conductivity of *n*-type system is lower than that of *p*-type system at the same carrier concentration, which is due to smaller relaxation time of the electrons compared with that of holes (see Table S1 of the Supplement Information). As the relaxation time changes little in the temperature range considered, we see from Fig. 3(d) that the variation of electrical conductivity with the carrier concentration at 800 K is very similar to that at 300 K. Combined with a relatively higher Seebeck coefficient at elevated temperature, we expect that the power factor of BP at 800 K is higher than that at 300K. For the electronic thermal conductivity $\kappa_e$, as the Lorenz number of *n*-type system calculated from Eq. (5) is roughly equal to that of *p*-type system at the same carrier concentration (see Supplement Information), it is reasonable to expect that $\kappa_e$ would show similar behavior as $\sigma$ and is thus not shown here.

**3.3 Phonon transport properties**

We next discuss the phonon transport properties of BP. Within the phonon Boltzmann transport theory, the lattice thermal conductivity of BP can be calculated without adjustable parameters. Figure 4(a) plots the phonon dispersion relations of BP, which is consistent with previous result using bond charge model [44]. Among the three acoustic phonon modes, we see the phonon group velocity along Γ-A direction (*x* direction or the zigzag direction) is the highest, while it is the lowest along the Γ-Z direction (*y* direction or the out-of-plane direction), and that along Γ-Y direction (*z* direction or the armchair direction) is in between [44]. According to Eq. (6), we expect that the lattice thermal conductivity is the highest along the *x* direction, followed by that along the *z* direction and then along the *y* direction. This is indeed the case as shown in Figure 4(b), where the lattice thermal conductivity of BP is plotted as a function of temperature in the range from 300 K to 800 K. The reason that the thermal conductivity along the *y* direction is the lowest can be attributed to the fact that the P atoms in the out-of-plane direction are connected by van der Waals force while it is covalently bonded in the plane. At 800 K, the calculated thermal conductivity of BP is 57.5, 24.1 and 22.7 W/mK along the *x*, *z*, and *y* direction, respectively. It seems that such large thermal conductivity may be not suitable for thermoelectric applications if compared with that of conventional thermoelectric materials. However, we will see in the following that the thermoelectric performance of BP is still considerable along the *z* direction due to relatively higher power factor.

**3.4 Thermoelectric figure of merit**

With all the transport coefficients available to us, we can now estimate the *ZT* value of BP. Figure 5(a) and 5(b) show the *ZT* values as a function of carrier concentration at 300 K and 800 K, respectively. In either case, the maximum *ZT* is achieved in the *n*-type system and found to be that along the *z* direction (the armchair direction). We see that at room temperature, the optimal *ZT* value is very small (0.06) caused by a large lattice thermal conductivity (65.1 W/mK). Note that such a *ZT* is actually lower than that predicted by Qin *et al.* [9] where an experimentally measured lattice thermal conductivity of polycrystalline BP is used (12.1 W/mK) and is obviously lower than our calculated result. When the temperature is increased to 800 K, there is a significant reduction of the thermal conductivity (from 65.1 W/mK to 24.1 W/mK). As a result, we see from the Fig. 5(b) that the BP exhibits the largest *n*-type *ZT* value of 1.1 at a carrier concentration of $1.5 \times 10^{20}$ cm$^{-3}$, and *p*-type *ZT* value of 0.6 at a carrier concentration of $3.2 \times 10^{19}$ cm$^{-3}$. As mentioned before, although the lattice thermal conductivity is still large at such temperature, the thermoelectric efficiency is considerable since the optimal power factor can reach 0.064 W/mK$^2$, which is higher than those of state-of-the-art thermoelectrics [45, 46]. Our theoretical calculations thus indicate that BP could be an intermediate temperature thermoelectric material by properly controlling the carrier concentration. In Fig. 5 (c), we plot the temperature dependence of the optimal *ZT* values. We see that the *ZT* increase monotonically with the temperature, which is caused by the increase of Seebeck coefficient and decrease of lattice thermal conductivity. In general, the *n*-type BP exhibits larger *ZT* value than *p*-type system. Although the lattice thermal conductivity along the out-of-plane direction is the lowest among the three directions, the thermoelectric performance along this direction is poor due to relatively lower electrical conductivity (see Fig. 3(c) and Fig. 3(d)).

**3.5 Further optimization of *ZT* value**

Up to now, we have systematically investigated the thermoelectric properties of pristine BP. We find that at room temperature, the thermoelectric efficiency of BP is poor due to relatively larger lattice thermal conductivity. Even one can obtain an optimal *ZT* of 1.1 at 800 K, it is still less than those of the best thermoelectric materials. To significantly enhance the thermoelectric performance, an efficient approach is by doping an element with similar electronic configuration but different atomic mass to form a solid solution, which can lower the thermal conductivity by inducing local distortion without seriously affecting the electrical conductivity [47]. As an example, we show below the effect of substitution of P atom with Sb atom on the thermoelectric properties of BP.

We construct our model by replacing a P atom with a Sb atom in the unit cell, which gives a nominal formula of $P_{0.75}Sb_{0.25}$. The phonon spectrum of $P_{0.75}Sb_{0.25}$ is given in the Supplement Information, which shows no imaginary frequency and suggests the stability of the substituted system. The optimized lattice constant of $P_{0.75}Sb_{0.25}$ is *a*=3.66 Å, *b*=11.12 Å, and *c*=4.48 Å, which is larger than that of pristine BP due to larger radius of Sb atom. The calculated electronic band structure and the corresponding DOS are shown in Figure 6. We see from Fig. 6(a) that $P_{0.75}Sb_{0.25}$ is also a direct band gap semiconductor with the band edges located at the Z point. More interestingly, there are several energy pockets around the Fermi level with almost the same energy as that of the band extrema. For example, the second low energy pocket (indicated by the red circle) appears at the X point and has an energy of 51.4 meV above the conduction band minimum (CBM). The second high energy pocket (indicated by the blue circle) appears at the $\Gamma$-Y direction and has an energy of 19.1 meV lower than that of the valence band maximum (VBM). This finding implies that Sb substitution could be used to increase the DOS of BP by inducing additional energy pockets around the Fermi level, which is very beneficial to their thermoelectric performance. Indeed, we find from Fig. 6(b) that the DOS at the band edge for the substituted system become steeper than that of pristine BP, thus leading to an increased Seebeck coefficient of $P_{0.75}Sb_{0.25}$ (see Figure S4 of the Supplement Information).

As the thermoelectric performance of *n*-type BP is the best along the armchair direction, we focus on such direction for the *n*-type $P_{0.75}Sb_{0.25}$. Figure 7(a) plots the lattice thermal conductivity of $P_{0.75}Sb_{0.25}$ as a function of temperature from 300 K to 800 K. For comparison, the result for pristine BP is also shown. We see that substituting P with Sb leads to dramatic reduction of lattice thermal conductivity in the whole temperature range considered. At room temperature, the thermal conductivity of $P_{0.75}Sb_{0.25}$ is calculated to be 7.3 W/mK, which is 89% smaller than that of pristine BP. Moreover, this value can be further reduced to 2.7 W/mK at 800 K. Summarizing these results, we find that Sb substitution can not only enhance the electronic transport of BP by increasing the DOS around the Fermi level, but also suppress the lattice thermal conductivity because of the mass difference between Sb and P atom. It is thus reasonable to expect that the thermoelectric performance of $P_{0.75}Sb_{0.25}$ could be dramatically enhanced. Indeed, we see from Figure 7(b) that the *ZT* value of the $P_{0.75}Sb_{0.25}$ can reach 0.78 at 300 K, which is much larger than that of pristine BP (*ZT*=0.06). Moreover, the *ZT* values of $P_{0.75}Sb_{0.25}$ exhibit strong temperature dependence and can be optimized to as high as 5.4 at 800 K, which is significantly larger than that of pristine BP (*ZT*=1.1). We further find that the *ZT* value of the Sb-substituted system is larger than 2.0 at relatively broad temperature range from 450 K to 800 K, which makes $P_{0.75}Sb_{0.25}$ a very promising candidate for thermoelectric applications if operated at an intermediate temperature region.

## 4. Summary

We demonstrate by first-principles calculations and Boltzmann theory that at high temperature, the BP exhibits considerable thermoelectric performance along the armchair direction, even with a relatively higher lattice thermal conductivity. Besides, we show that Sb substitution can not only improve the electronic transport by increasing the DOS around the Fermi level, but also suppress the lattice thermal transport caused by the mass difference between Sb and P atom. As a result, the optimal *ZT* value of the substituted system ($P_{0.75}Sb_{0.25}$) can reach as high as 5.4 at 800 K. To experimentally realize such goals, one needs to first fabricate the Sb-substituted BP. Near the completion of this work, we become aware of the black arsenic-phosphorus (b-$As_xP_{1-x}$), a layered structure similar to $P_{0.75}Sb_{0.25}$ is rationally synthesized with highly tunable chemical compositions [48, 49]. Since Sb atom has a similar electronic configuration with As atom, we believe that our investigated $P_{0.75}Sb_{0.25}$ can be produced in a similar way. With the rapid developments of fabrication techniques, we anticipate that the present work will stimulate new experimental efforts and advances in the search of high performance thermoelectric materials.

**Acknowledgments**

We thank financial support from the National Natural Science Foundation (Grant No. 51172167 and J1210061) and the "973 Program" of China (Grant No. 2013CB632502).

**Table I** The lattice constant ($a$, $b$, $c$), bond distance ($d_1$, $d_2$), bond angles ($\theta_1$, $\theta_2$), and interlayer van der Waals distance ($d_0$) obtained by using different exchange and/or correlation functionals. The corresponding experiment values are also shown for comparison.

| functionals | $a$ (Å) | $b$ (Å) | $c$ (Å) | $d_1$ (Å) | $d_2$ (Å) | $\theta_1$ (°) | $\theta_2$ (°) | $d_0$ (Å) |
|---|---|---|---|---|---|---|---|---|
| LDA | 3.26 | 10.22 | 4.20 | 2.20 | 2.24 | 95.44 | 100.67 | 2.95 |
| PBE | 3.34 | 11.14 | 4.52 | 2.23 | 2.26 | 96.79 | 103.21 | 3.45 |
| vdW-DF | 3.42 | 11.34 | 4.63 | 2.27 | 2.29 | 97.60 | 103.60 | 3.53 |
| vdW-DF2 | 3.48 | 11.30 | 4.66 | 2.30 | 2.31 | 98.35 | 103.51 | 3.49 |
| optPBE-vdW | 3.38 | 10.90 | 4.50 | 2.26 | 2.28 | 97.02 | 102.69 | 3.30 |
| optB88-vdW | 3.36 | 10.70 | 4.45 | 2.25 | 2.28 | 96.58 | 102.27 | 3.19 |
| optB86b-vdW | 3.32 | 10.54 | 4.36 | 2.24 | 2.27 | 95.90 | 101.63 | 3.11 |
| DFT-D2 | 3.32 | 10.46 | 4.42 | 2.23 | 2.26 | 96.57 | 102.36 | 3.09 |
| Exp. | 3.31 | 10.48 | 4.38 | 2.22 | 2.24 | 96.34 | 102.09 | 3.07 |

**Table II** Calculated carrier effective mass of BP (in unit of inertial mass of electron $m_0$) along the $x$, $y$, and $z$ directions.

| carrier type | $m_x^*$ ($m_0$) | $m_y^*$ ($m_0$) | $m_z^*$ ($m_0$) |
|---|---|---|---|
| hole | 0.833 | 0.335 | 0.144 |
| electron | 1.268 | 0.164 | 0.163 |

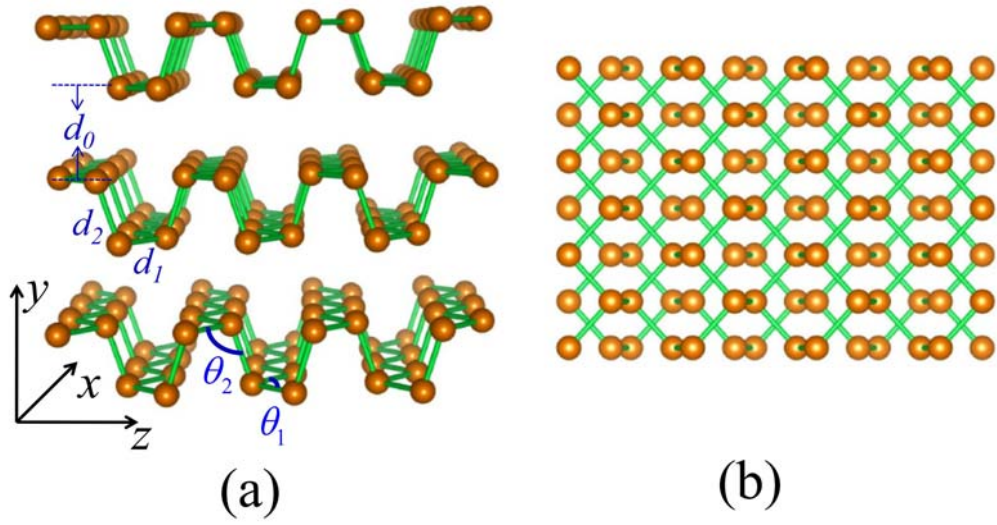

**Figure 1** The ball-and-stick model of BP: (a) side-view, and (b) top-view. The coordinate axes ($x$, $y$, $z$) and structural parameters ($d_0$, $d_1$, $d_2$, $\theta_1$, $\theta_2$) are indicated.

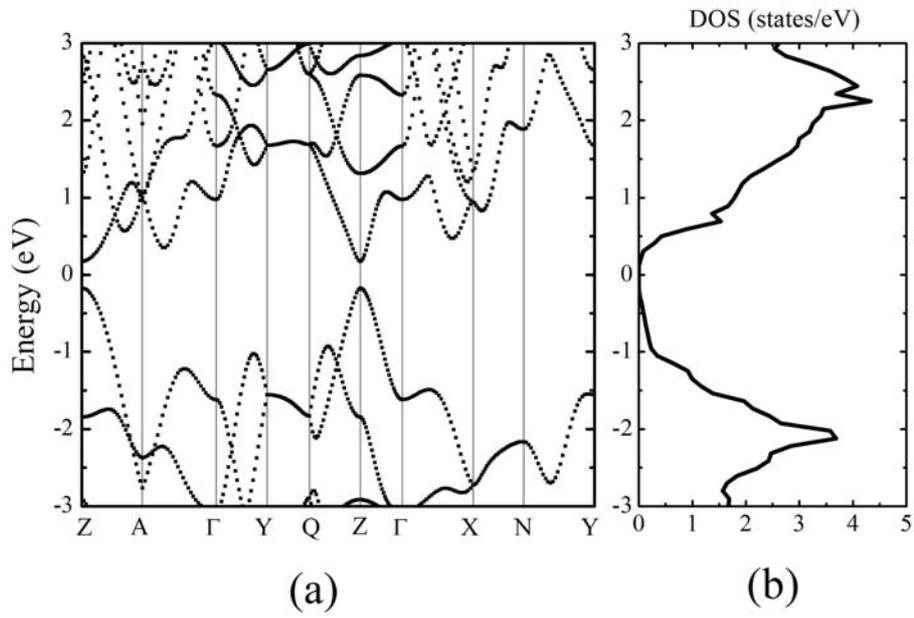

**Figure 2** (a) The calculated band structure of BP, and (b) the corresponding DOS.

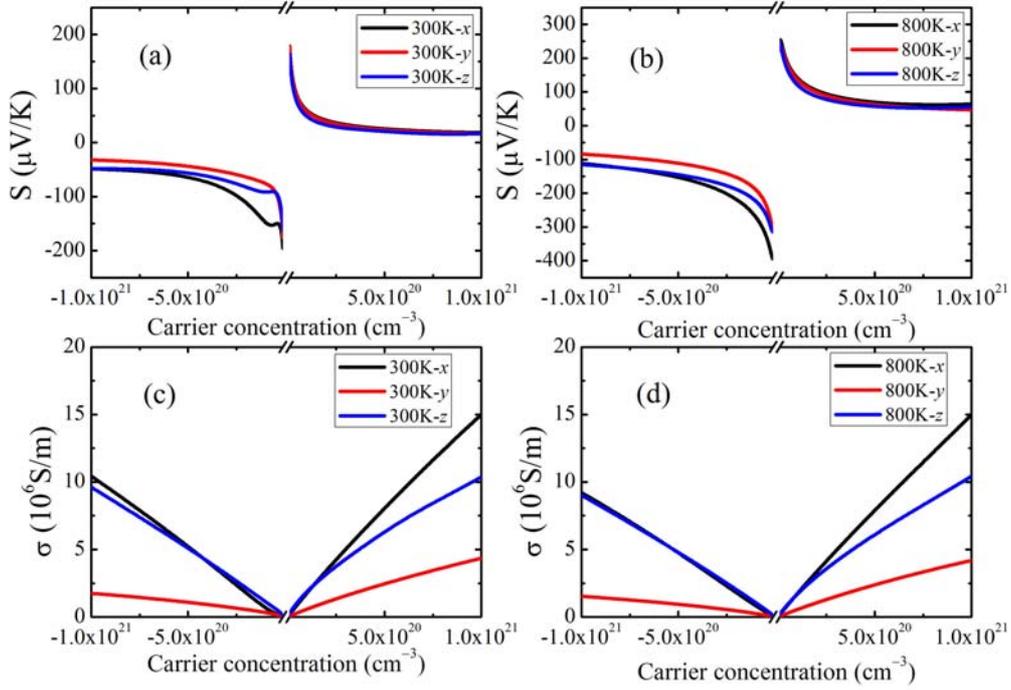

**Figure 3** The electronic transport coefficients of BP as a function of carrier concentration: (a) and (b) are the Seebeck coefficient at 300 K and 800 K, respectively. (c) and (d) are the electrical conductivity at 300 K and 800 K, respectively. Negative and positive carrier concentrations represent *n*- and *p*- type carriers, respectively.

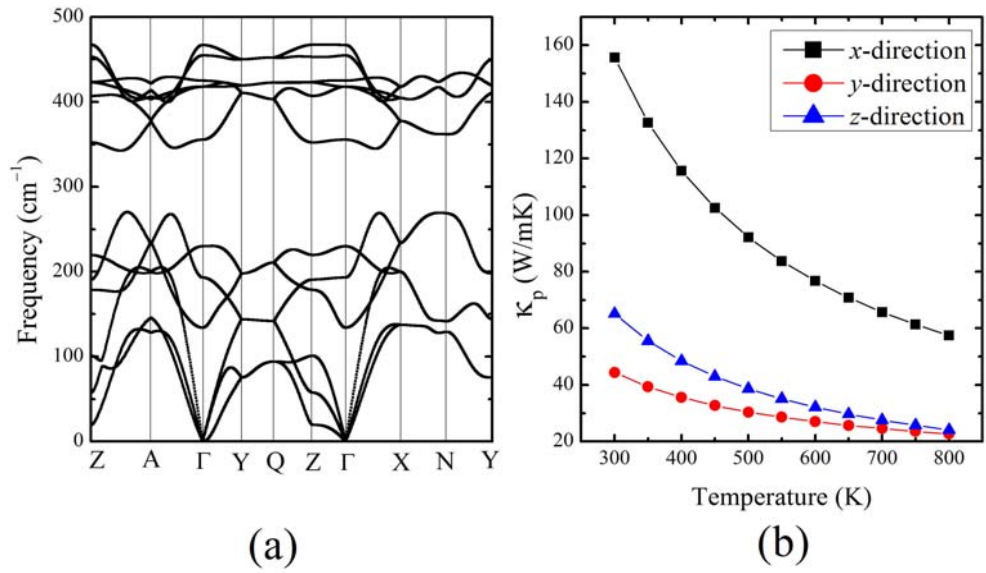

**Figure 4** (a) The phonon dispersion relations of BP, and (b) the calculated lattice thermal conductivity of BP as a function of temperature along three directions.

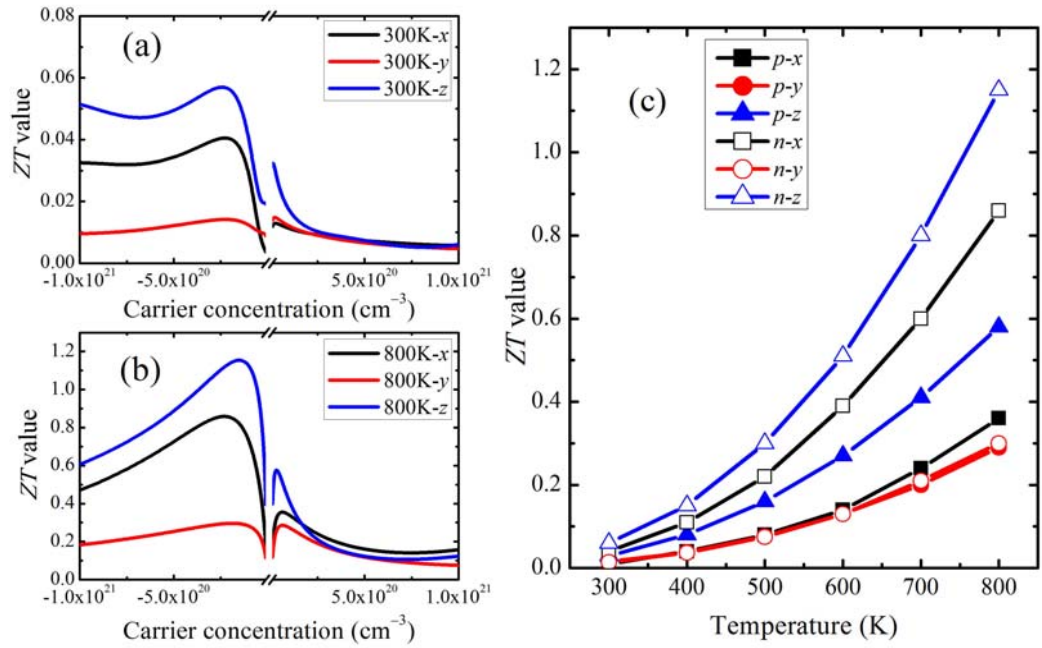

**Figure 5** The calculated *ZT* values of BP as a function of carrier concentration at (a) 300 K, and (b) 800 K. (c) is the optimal *ZT* values of BP as a function of temperature.

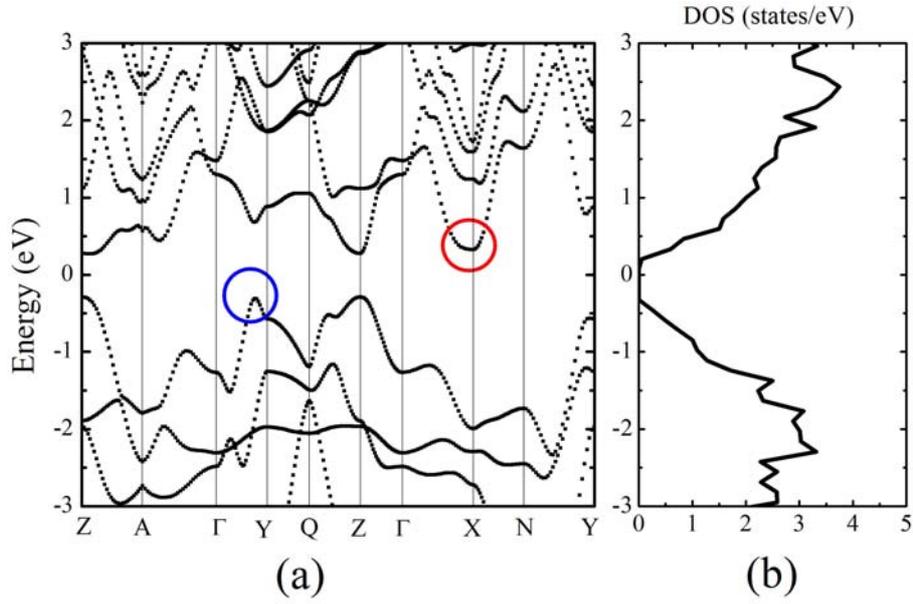

**Figure 6** (a) The calculated band structure of $P_{0.75}Sb_{0.25}$, and (b) the corresponding DOS. The red and blue circles indicate the second low and high energy pockets, respectively.

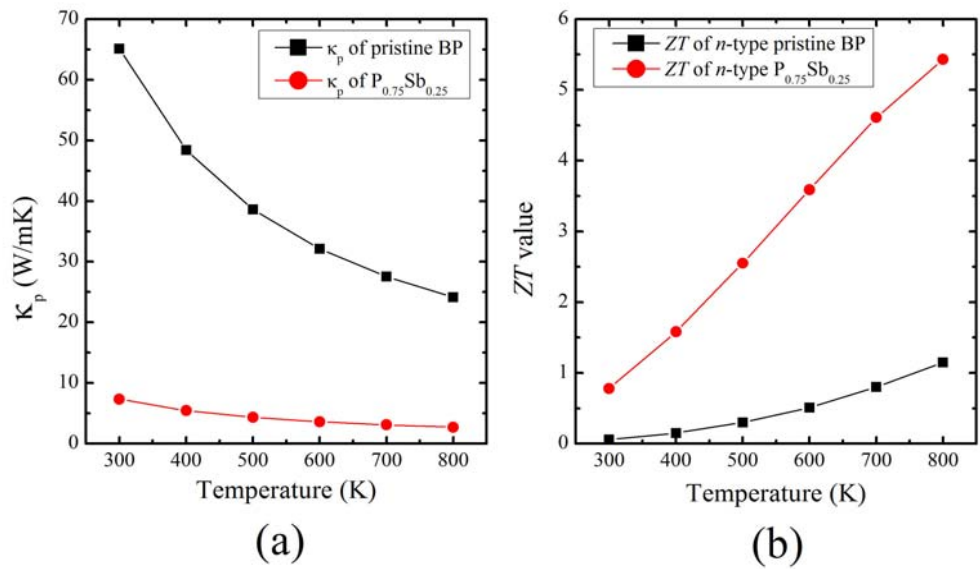

**Figure 7** Comparison of temperature dependence of (a) lattice thermal conductivity, and (b) *ZT* values of the *n*-type BP and P$_{0.75}$Sb$_{0.25}$ along armchair direction.

# References

[1] K. F. Hsu, S. Loo, F. Guo, W. Chen, J. S. Dyck, C. Uher, T. Hogan, E. K. Polychroniadis, and M. G. Kanatzidis, Science **303**, 818 (2004).

[2] B. Poudel, Q. Hao, Y. Ma, Y. Lan, A. Minnich, B. Yu, X. Yan, D. Wang, A. Muto, D. Vashaee, X. Chen, J. Liu, M. S. Dresselhaus, G. Chen, and Z. Ren, Science **320**, 634 (2008).

[3] K. Biswas, J. He, I. D. Blum, C.-I Wu, T. P. Hogan, D. N. Seidman, V. P. Dravid, and M. G. Kanatzidis, Nature **489**, 414 (2012).

[4] R. Venkatasubramanian, E. Siivola, T. Colpitts, and B. O'Quinn, Nature **413**, 597 (2001).

[5] T. C. Harman, P. J. Taylor, M. P. Walsh, and B. E. LaForge, Science **297**, 2229 (2002).

[6] A. I. Hochbaum, R. K. Chen, R. D. Delgado, W. J. Liang, E. C. Garnett, M. Najarian, A. Majumdar, and P. D. Yang, Nature **451**, 163 (2008).

[7] H. Y. Lv, W. J. Lu, D. F. Shao, and Y. P. Sun, arXiv: 1404.5171 (2014).

[8] G. A. Slack, Phys. Rev. **139**, 507 (1965).

[9] G. Z. Qin, Q.-B. Yan , Z. Z. Qin, S.-Y. Yue, H.-J. Cui, Q.-R. Zheng, and G. Su, Sci. Rep. **4**, 6946 (2014).

[10] E. Flores, J. R. Ares, A. Castellanos-Gomez, M. Barawi, I. J. Ferrer, and C. Sánchez, Appl. Phys. Lett. **106**, 022102 (2015).

[11] G. Kresse and J. Hafner, Phys. Rev. B **47**, R558 (1993).

[12] G. Kresse and J. Hafner, Phys. Rev. B **49**, 14251 (1994).

[13] G. Kresse and J. Furthmüller, Comput. Mater. Sci. **6**, 15 (1996).

[14] G. Kresse and J. Furthmüller, Phys. Rev. B **54**, 11169 (1996).

[15] J. P. Perdew, K. Burke, and M. Ernzerhof, Phys. Rev. Lett. **77**, 3865 (1996).

[16] H. J. Monkhorst and J. D. Pack, Phys. Rev. B **13**, 5188 (1976).

[17] J. Klimeš, D. R. Bowler, and A. Michaelides, Phys. Rev. B **83**, 195131 (2011).

[18] J. Klimeš, D. R. Bowler, and A. Michaelides, J. Phys.: Condens. Matter **22**, 022201 (2010).

[19] A. D. Becke and E. R. Johnson, J. Chem. Phys. **124**, 221101 (2006).

[20] F. Tran and P. Blaha, Phys. Rev. Lett. **102**, 226401 (2009).

[21] G. K.H. Madsen and D. J. Singh, Comput. Phys. Commun. **175**, 67 (2006).

[22] D. J. Singh and I. I. Mazin, Phys. Rev. B **56**, R1650 (1997).

[23] G. K.H. Madsen, K. Schwarz, P. Blaha, and D. J. Singh, Phys. Rev. B **68**, 125212 (2003).

[24] D. J. Singh, Phys. Rev. B **81**, 195217 (2010).

[25] D. Parker and D. J. Singh, Phys. Rev. B **85**, 125209 (2012).

[26] A. Morita, Appl. Phys. A **39**, 227 (1986).

[27] A. Bejan and A. D. Allan, *Heat Transfer Handbook* (Wiley, New York, 2003), p. 1338.

[28] L. D. Hicks and M. S. Dresselhaus, Phys. Rev. B **47**, 12727 (1993).

[29] L. D. Hicks and M. S. Dresselhaus, Phys. Rev. B **47**, 16631 (1993).

[30] W. Li, J. Carrete, N. A. Katcho, and N. Mingo, Comput. Phys. Commun. **185**, 1747 (2014).

[31] W. Li, N. Mingo, L. Lindsay, D. A. Broido, D. A. Stewart, and N. A. Katcho, Phys. Rev. B **85**, 195436 (2012).

[32] W. Li, L. Lindsay, D. A. Broido, D. A. Stewart, and N. Mingo, Phys. Rev. B **86**, 174307 (2012).

[33] G. Z. Qin, Q.-B. Yan, Z. Z. Qin, S.-Y. Yue, M. Hu, and G. Su, Phys. Chem. Chem. Phys. **17**, 4854 (2015).

[34] L. Y. Zhu, G. Zhang, and B. W. Li, Phys. Rev. B **90**, 214302 (2014).

[35] A. Jain and A. J. H. McGaughey, Sci. Rep. **5**, 8501 (2015).

[36] J. Zhang, H. J. Liu, L. Cheng, J. Wei, J. H. Liang, D. D. Fan, J. Shi, X. F. Tang, and Q. J. Zhang, Sci. Rep. **4**, 6452 (2014).

[37] S. Grimme, J. Comput. Chem. **27**, 1787 (2006).

[38] A. Brown and S. Rundqvist, Acta Cryst. **19**, 684 (1965).

[39] J. S. Qiao, X. H. Kong, Z.-X. Hu, F. Yang, and W. Ji, Nat. Commun. **5**, 4475 (2014).

[40] R. W. Keyes, Phys. Rev. **92**, 580 (1953).

[41] D. Warschauer, J. Appl. Phys. **34**, 1853 (1963).

[42] Y. Akahama, S. Endo, and S. Narita, J. Phys. Soc. Jpn. **52**, 2148 (1983).

[43] J. M. Yang, G. Yang, G. B. Zhang, and Y. X. Wang, J. Mater. Chem. A **2**, 13923 (2014).

[44] C. Kaneta, H. K. Yoshida, and A. Morita, J. Phys. Soc. Jpn. **55**, 1213 (1986).

[45] G. J. Snyder and E. S. Toberer, Nature Mater. **7**, 105 (2008).

[46] L. D. Zhao, V. P. Dravid, and M. G. Kanatzidis, Energy Environ. Sci. **7**, 251

(2014).

[47] C. Wood, Rep. Prog. Phys. **51**, 459 (1988).

[48] B. Liu, M. Köpf, A. A. Abbas, X. Wang, Q. Guo, Y. Jia, F. Xia, R. Weihrich, F. Bachhuber, F. Pielnhofer, H. Wang, R. Dhall, S. B. Cronin, M. Ge, X. Fang, T. Nilges, and C. Zhou, arXiv: 1505.07061 (2015).

[49] O. Osters, T. Nilges, F. Bachhuber, F. Pielnhofer, R. Weihrich, M. Schöneich, and P. Schmidt, Angew. Chem. Int. Ed. **51**, 2994 (2012).